\documentclass[aps,pra,floats,epsfig,pdflatex]{revtex4}                                                              %

\usepackage{url}
\usepackage{amsbsy}
\usepackage{amsmath}
\usepackage{amssymb}
\usepackage{amsfonts}
\usepackage{graphics}
\usepackage[update,prepend]{epstopdf}
\usepackage{listings}
\usepackage{hyperref}
\usepackage{lineno}
\begin{document}


\title{Nonequilibrium Dynamics of Ultracold Fermi Superfluids.}

\author{Analabha Roy}                                                                                                %
\affiliation{TCMP division, Saha Institute of Nuclear Physics, $1/AF$ Bidhannagar, Kolkata-700064, India.}             %
\begin{abstract}
The aim of this mini review is to survey the literature on the study of nonequilibrium dynamics of Fermi 
superfluids in the BCS and BEC limits, both in the single channel and dual channel cases. The focus is on mean field 
approaches to the dynamics, with specific attention drawn to the dynamics of the Ginzburg-Landau order parameters of the 
Fermi and composite Bose fields, as well as on the microscopic dynamics of the quantum degrees of freedom. The two 
approaches are valid approximations in two different time scales of the ensuing dynamics. The system is presumed to 
evolve during and/or after a quantum quench in the parameter space. The quench can either be an impulse quench with 
virtually instantaneous variation, or a periodic variation between two values. The literature for the order parameter 
dynamics, described by the time-dependent Ginzburg- Landau equations, is reviewed, and the works of the author in this 
area highlighted. The mixed phase regime in the dual channel case is also considered, and the dual order parameter 
dynamics of Fermi-Bose mixtures reviewed. Finally, the nonequilibrium dynamics of the microscopic degrees of 
freedom for the superfluid is reviewed for the self-consistent and non self-consistent cases. The dynamics of the former 
can be described by the Bogoliubov de-Gennes equations with the equilibrium BCS gap equation continued in time and self
-consistently coupled to the BdG dynamics. The latter is a reduced BCS problem and can be mapped onto the dynamics of 
Ising and Kitaev models. This article reviews the dynamics of both impulse quenches in the Feshbach detuning, as well as 
periodic quenches in the chemical potential, and highlights the author's contributions in this area of research.
\end{abstract}                                                                                                       %

\maketitle
\section{Introduction}
\label{intro}
 Studies of the coherent quantum dynamics of closed many body systems have gained considerable interest in the last two decades. This interest has been brought about by such states becoming experimentally accessible using ultracold atoms, cooled to near absolute zero temperatures by the use of optical molasses~\cite{mollases:seminal} , evaporative cooling~\cite{bec},  laser culling~\cite{culling} , cavity QED methods~\cite{mmc} , and other experimental techniques. The experimental realization of the Bose Einstein condensate by Cornell, Ketterle and Weiman in $1995$~\cite{bec} , as well as the realization of the Bosonic Superfluid Mott-Insulator transition by Greiner \textit{et. al.} in $2002$~\cite{greiner} (theoretically predicted by Jaksch \textit{et. al.} in $1998$~\cite{jaksch} ), motivated strong theoretical interest in the coherent dynamics of ultracold bosons. Previously, the regimes where fermionic coherent dynamics can be realized were unattainable in either solid state systems or with ultracold 
atoms. The difficulties for ultracold atoms were eventually circumvented by experiments performed over the first decade of the $21^{\rm st}$ century. Studies of fermionic atoms gained momentum after these experiments realized  pure fermionic BCS states~\cite{exp1}  , as well as the BCS-BEC crossover that was predicted theoretically a decade earlier~\cite{bcsbec} . The formation of Fermi superfluids were realized by the use of strong four fermion interactions caused by tuning a homogeneous magnetic field close to the Feshbach resonance of the confined atoms~\cite{feshbach}. Such strong interactions are difficult to generate with bosonic atoms due to three body losses. However, three body losses are significantly lower with fermions due to the Pauli principle. Theoretical interest in Fermi superfluid dynamics paralleled experiments involving dynamical phenomena. Theoretical studies of the exact dynamics focused on BCS-
BEC systems near unitarity or in the deep BCS regime. However, the primary theoretical interest continues to remain with Bosonic dynamics, particularly the Bose-Hubbard model, and Ising-like spin models.

This mini review focuses on theoretical works investigating the other end of the spectrum \textit{i.e}, the dynamics of Fermi superfluids. The exclusive focus is on mean field dynamics at $T=0$ as the system is quenched in various ways. The literature can be broadly divided into two kinds, one that looks at the macroscopic dynamics of the order parameters, covered in section~\ref{sec:myoldstuff},  {and one that looks at the dynamics of the microscopic degrees of freedom, covered in section}~\ref{sec:fermidyn}.  {In the former case, the excited quasiparticles in the Bogoliubov de-Gennes spectrum of the superfluid relax very quickly compared to the relaxation of the time dependent superfluid gap $\Delta(t)$, and excitations of the quasimolecular states relax similarly}~\cite{myrefs, huang:becbcs2}. Thus, the dynamics of the system can be described by a mean field treatment of the macroscopic order parameter $\Delta(t)$ in the single channel Feshbach resonance, and a two-parameter  system spanned by $\Delta(t)$ 
and the BEC matter field $b_0(t)$ in the dual channel case. In the latter case, the Bogoliubov excitations relax much more slowly, and the system must be described by the dynamics of the quasiparticles with a coupling to a self-consistent equation for $\Delta(t)$. These regimes are discussed in more detail in~\cite{abrahams,barankov, andreev:noneqmbcsbec, gorkov, volkov}, as well as in~\cite{aronov,lo}.  {This review restricts attention to the mean field dynamics of population balanced Fermi gases at $T=0$ for Fermi superfluids that can be modeled by integrable closed quantum systems. The dynamics of fluctuations, nonintegrable systems, and systems connected to thermal reservoirs, involve diagrammatic techniques} (see, for instance, ref~\cite{gorkov}) , time dependent Density Matrix Renormalization Group techniques (see, for instance,~\cite{dmrg:review}), and the Eigenstate Thermalization Hypothesis (see, for instance, ref~\cite{krishrev})  { respectively. These are beyond 
the scope of this review, and the reader is directed to the references cited above for details.

This mini-review article contains results from}~\cite{abrahams,barankov,arnab1,mypaper4,ourpaper}. Section~\ref{sec:fermidyn} contains material from~\cite{ourpaper} that was presented by the author at the $2012$ National Conference on Nonlinear Systems and Dynamics (Quantum Chaos minisymposium), held at the Indian Institute of Science Education and Research (IISER), Pune, India.  { Its new contribution consists in generalizing}~\cite{barankov} {to explicitly time dependent quenches, as well as suggesting future directions in this area of research based on }~\cite{volkov,barankov,ourpaper} {. This mini review also connects the different models and paradigms used in the mean field dynamics of Fermi BCS systems, and highlights literature in an area of research that has gained considerable interest in the last few years.}

\section{Dynamics of the Order Parameters}
\label{sec:myoldstuff}
 {This section details studies of the dynamics of the macroscopic order parameters that arise from the microscopic BCS-BEC Hamiltonian. Studies of these regimes have been motivated by the BCS-BEC crossover, where the renormalized interaction between the Cooper pairs changes sign, and the size of the Cooper pairs decrease. These pairs eventually become diatomic molecules that obey Bose statistics and can form a Bose Einstein Condensate. The equilibrium equation of state of the macroscopic order parameters has also shown the BCS-BEC crossover}~\cite{huang:becbcs1} {, and this section details works that involve the nonequilibrium extension of these equations.} 

 {The dynamics of ultracold Fermi superfluids can be understood by approximate models based on how the time scales of the dynamics compare to the relaxation of the quasiparticles of the equilibrium BCS Hamiltonian. The order parameter relaxes at a rate given by $\hbar \Delta^{-1}$, where $\Delta$ is the macroscopic BCS energy gap responsible for superconducting/superfluid behavior. When this relaxes much more slowly than the energies of the quasiparticles, they reach their equilibrium quickly and the nonequilibrium state can be characterized by the macroscopic order parameters. This, however, only occurs in very specific regions in the full parameter space}~\cite{gorkov}  {near the critical temperature. More general mean field treatments are discussed in section}~\ref{sec:fermidyn} {. Nonetheless, recent works have focused on investigating the order parameter dynamics of Fermi Bose mixtures. The most general \textit{dual channel model} has been espoused by Holland \textit{et. al.}}~\cite{holland} {and 
Timmermans \textit{et. al}}~\cite{timmermans}  {as a model for BCS superfluidity of neutral fermionic atoms that interact attractively via a Feshbach resonance that is deep enough to induce molecular bonding.} Most recently, the ensuing dynamics of mean field order parameters following an impulse quench of a system parameter at $t=0-$ has also been studied. (Roy~\cite{mypaper4}).

 {The dynamical equations of motion can be obtained from the $D-$dimensional \textit{Timmermans' Hamiltonian}}~\cite{timmermans}  {$H_{tm}$ for a Fermi-Bose mixture at $T=0$ confined in a unit volume. Large impulse quenches in the Feshbach detuning can produce persistent oscillations in the matter wave field due to the nonlinear modes, similar to the collapse and revival of matter fields in such systems reported in the literature}~\cite{colrev,huang:becbcs1} {. The \textit{Timmermans' Hamiltonian} in momentum space can be expressed in terms of $a_{{\bf k},\sigma}$, the annihilation operator for the BCS fermions for momentum ${\bf k}$  and (pseudo) spin $\sigma = \uparrow,\downarrow$, and $b_{\bf q}$, the operators of the composite bosons that are formed by the Feshbach attraction of two fermions of momenta $\pm {\bf p}+{\bf q}/2$ and opposite spin.}  {This yields}~\cite{mypaper4}
\begin{multline}
\label{eq:timmermans:kspace}
H_{tm} = \sum_{{\bf k},\sigma} \left(\epsilon_{\bf k} - \mu_F \right)a^\dagger_{{\bf k}\sigma}a^{ }_{{\bf k}\sigma} - |u_F|\sum_{{\bf k}{\bf k'}}a^\dagger_{{\bf k}\uparrow}a^\dagger_{-{\bf k}\downarrow}a^{ }_{{\bf k'}\downarrow}a^{ }_{{\bf k'}\uparrow}\\
+\sum_{\bf q} \left(E^0_{\bf q}+2\nu-\mu_B \right)b^\dagger_{\bf q}b^{ }_{\bf q} + u_B\sum_{{\bf q_1}{\bf q'_1} {\bf q_2} {\bf q'_2}}b^\dagger_{\bf q'_1}b^\dagger_{\bf q'_2} b_{\bf q_2} b_{\bf q_1}\delta_{{\bf q_1}+{\bf q_2},{\bf q'_1}+{\bf q'_2}}\\
+g_r\sum_{{\bf k}{\bf q}}\left(b^\dagger_{\bf q}a_{{\bf p+\frac{q}{2}}\uparrow}a_{{\bf -p+\frac{q}{2}}\downarrow} + h.c. \right).
\end{multline}
 {Here, the first line represents the fermions in momentum space with the BCS four-fermion interaction. The free fermion energies are represented by $\epsilon_{\bf k}$. The second line represents the Hamiltonian for the closed channel quasimolecular bosons $b^\dagger_{\bf q}$ ($b^{ }_{\bf q}$) with energies $E^0_{\bf q}+2\nu-\mu_B$ and point contact repulsion of amplitude $u_B$. Finally, the last line represents the atom-molecule coupling, through which the Fermi and Bose condensates exchange momentum ${\bf q}$. Huang \textit{et. al.}}~\cite{huang:becbcs2}  {have obtained the excitation spectra of this Hamiltonian in the mean field}, and Roy~\cite{mypaper4} has demonstrated that the relaxation time of these excitations are faster than $E^{-1}_c$, where $E_c = 2\sqrt{\mu^2_F + |\Delta + g_rb_0|^2}$, $\Delta$ is the equilibrium BCS gap computed from the gap equation~\cite{coleman}, and $b_0$ is the ground state population of the bosons. Therefore, impulse quenches that are slower than $E^{-1}_c$ allow an 
analysis of the dynamical system based only on a ground state BEC field and fermion pair field to completely capture the non-equilibrium state. The Timmermans Hamiltonian can be rewritten in coordinate space as follows.

\begin{equation}
 H_{tm} = \int{{\mathrm d}^Dx} \times {\mathcal H_{tm}} (x),
\end{equation}
where
\begin{multline}
\label{timmermans}
{\mathcal H_{tm}}(x) =  \sum_{\sigma} \bigg\{ \phi^\dagger_\sigma(x) \left[h_F({\bf r})-\mu_F\right]\phi_\sigma(x) \bigg \} - |u_F| \phi^\dagger_\uparrow(x)\phi^\dagger_\downarrow(x)\phi^{}_\downarrow(x)\phi^{}_\uparrow(x)\\ 
+b^\dagger(x)\left[h_B({\bf r})+2\nu-\mu_B \right]b^{ }(x)+ u_B b^\dagger(x) b(x) \left[b^\dagger(x) b(x)-1\right]\\
+g_r\left[b^\dagger(x) \phi_\uparrow(x) \phi_\downarrow(x) + h.c\right],
\end{multline}
the interactions of the excited quasimolecules have been ignored, and $x=({\bf r},t')$. The fermion and composite boson fields in this system are represented by the operators  $\phi_\sigma(x)$ and $b(x)$  respectively. Equation~\ref{timmermans} describes a system of ultracold electrically neutral two-component fermions interacting attractively. The first line in equation~\ref{timmermans} represents the \textit{Fermi-BCS} part of the Hamiltonian. Here, $h_{F}({\bf r})(h_{B}({\bf r}))=\left[ - \frac{\nabla^2}{2(4)m}+V({\bf r}) \right]$ are the single particle Hamiltonians for the fermions (Bosons), and $m$ is the fermion mass (the mass of the composite bosons is thus $2m$). The second line represents the Hamiltonian of the composite bosons~\cite{timmermans, huang:becbcs2}. Here, the Feshbach threshold energy (also called the Feshbach 'detuning' from the molecular channel to the continuum~\cite{timmermans}) is represented by $2\nu$, and $u_B$ represents the amplitude of the repulsion between the composite 
bosons. The final line describes the Feshbach resonance that leads to the fermions binding to (or 
dissociating from) the composite bosons, with  $g_r$ representing the atom-molecule coupling. Note that the chemical potentials satisfy $\mu_B=2\mu_F$. 

 {The mean field dynamics of the Fermi and Bose order parameters can be obtained from the Timmermans Hamiltonian via Ginzburg Landau Abrikosov Gor'kov theory, where the path integral grand partition function is transformed via the Hubbard Stratonovich method}\ ~\cite{hubbard:hstrans,coleman} {, introducing the fermion gap parameter $\Delta(t)$ via a Gaussian identity,}
\begin{equation}
\int {\mathcal D}[\Delta^*,\Delta] \exp{\left[-\int^\infty_0{{\mathrm d}x'}\frac{\Delta^*(x')\Delta(x')}{|u_F|}\right]}=1.
\end{equation}
 {Here, $ {\mathcal D}[\Delta^*,\Delta]$ is the path integral measure of the gap field. Performing the Hubbard-Stratonovich transformation, $\Delta\rightarrow \Delta + |u_F|\phi_\downarrow\phi_\uparrow $ and $\Delta^*\rightarrow \Delta^* + |u_F|\bar{\phi}_\uparrow\bar{\phi}_\downarrow$, cancels out the four-fermion term from $H_{tm}$. Now, integrating out the fermion fields using formal Grassman calculus allows for an effective action in terms of fermionic and bosonic order parameters. Machida and Koyama}~\cite{machida:dynamics}  {obtained the dynamical equations of the coupled Fermi and Bose fields using the method summarized above for the special case of free particles, where the excitations of order $E_c$ were taken into consideration. This yielded spatio-temporally varying fields given by the solutions to the following dynamical system.} 
\begin{eqnarray}
\label{timmermans:fulldyn}
i \frac{\partial}{\partial t}\Delta(x) &=&  a\Delta(x) + b g_r b(x)+ \frac{c}{4m}{\boldsymbol \nabla}^2\Delta(x)+\frac{g_r d}{4m}{\boldsymbol \nabla}^2b(x)\nonumber \\
                                       & & -f|\Delta(x)+g_r b(x)|^2 \left[\Delta(x)+g_r b(x)\right], \nonumber \\
i \frac{\partial}{\partial t}b(x) &=& -\frac{g_r}{|u_F|}\Delta(x)+\left(2\nu-\mu_B\right)b(x)-\frac{1}{4m}{\boldsymbol \nabla}^2 b(x).
\end{eqnarray}
 {Here, the complex constants $a-d$ and $f$ can be obtained from the perturbation amplitudes of the path integral action. Their values depend on the propagators of noninteracting and free Fermi and Bose gases. These equations can be easily generalized to trapped gases as well. The dynamics can also be obtained from the nonequilibrium Schwinger-Keldysh formalism by making the quasiparticle approximation described in the second paragraph of this section to the self-energy term formed by 1-particle- irreducible Feynman diagrams in the nonequilibrium Dyson equation~\cite{rammer}. The dynamics represents a coupled system of Ginzburg-Landau equations (for the Fermi field $\Delta(x)$) and Gross-Pitaevski equations (for the Bose field $b(x)$). If the atom molecule coupling $g_r$ is ignored, the dynamics of the Fermi and Bose fields decouple. This special case of
the dynamics in Eq.}~\ref{timmermans:fulldyn} { reproduces that of the \textit{single channel} Fermi-BCS model, where the Cooper pairs do not form bound states, and the Feshbach detuning is small enough to keep them in the scattering continuum. In such a regime, Eqs.}~\ref{timmermans:fulldyn} { decouple and the first equation becomes the time dependent Ginzburg Landau equation for the order parameter $\Delta(t)$.  The nonlinear damping in the dynamics produces very rich and diverse behavior in Fermi-Bose mixtures. The dynamics of the single parameter Ginzburg-Landau equation, a special kind of Nonlinear Schr\"odinger equation, has been investigated thoroughly over several decades in various contexts ranging from Poiseuille flows}~\cite{pflowgl} {, reaction-diffusion systems}~\cite{rdgl} {, as well as Fermi BCS superfluids and BEC's}~\cite{glreview} {. Stable solitonic solutions have been identified in certain regimes, as well as instabilities in lower dimensions obtained using various criteria} (
for 
instance, Derrick's theorem~\cite{derrick,khare} or the Vakhitov and Kolokolov stability criteria~\cite{vakhitov,khare} for $U(1)$ invariant Hamiltonians) {. These will not be reviewed here, and the reader is directed to ref}~\cite{glreview} {, and references therein, for details.}

General solutions to the dynamics in Eqs.~\ref{timmermans:fulldyn} have been discussed by Machida and Koyama~\cite{machida:dynamics}, as well as 
Chen and Guo~\cite{weaksols}, Guo, Fang and Wang~\cite{travellingsols}, and others. The rest of this section focuses on the special case when quasiparticle excitations can be ignored. In that limit, the fermion field (characterized by the gap parameter) is spatially homogeneous due to it's coupling  to the spatially homogeneous Bose field $b_0$ by momentum conservation. Then, the expansion of the effective action to the fourth order and subsequent computation of the functional derivatives with respect to the fields yields the mean field dynamics of the order parameters following an analytic continuation of time to the real axis. Neglecting the excitations away from the BCS state of composite bosons in Eqs.~\ref{timmermans:fulldyn}, and projecting the dynamics into the space spanned by the BEC and fermion pair field, yields the simplified dynamical system,
\begin{eqnarray}
\label{dynamical_system}
\dot \Psi_1 + i\gamma \left( \Psi_1-\Psi_2 \right) - i\alpha \Psi_1 +i\beta |\Psi_1|^2 \Psi_1 &=& 0 \nonumber \\
\dot \Psi_2 +2 i \lambda \Psi_2 +2 i\chi|\Psi_2|^2\Psi_2 - i\kappa\gamma \left(\Psi_1-\Psi_2 \right) &=& 0,
\end{eqnarray}
Here, $\Psi_i(t)=\big[\Delta(t)\delta_{i,1}+g_r b_0(t)\big]/\left(|\mu_F|\mathcal{N}\right)$, are the order parameter fields that describe the dynamics of the fermions in the BCS state and zero momentum bosons in the BEC respectively, and $\mathcal{N}$ is the total (fermion) particle number. The constants expressed as Greek letters in equation~\ref{dynamical_system} are evaluated from the perturbation amplitudes of the path integral. Their values depend on the propagators of a noninteracting Fermi gas in the trapping potential. These constants are complex and have been evaluated by Machida and Koyama~\cite{machida:dynamics}, and specific expressions for particle in a box confinement have been obtained by Roy~\cite{mypaper4}. 

The specific case of the particle in a box trap for the dual channel case,  as covered by Roy~\cite{mypaper4}, is of particular interest, and can be easily generalized to other confinements. The equilibrium fixed points of the dynamical system in Eq.~\ref{dynamical_system}, together with the imposition of number conservation, $\mathcal{N}=\lim_{\beta\rightarrow\infty}\partial_{\mu_F}\ln Z(\beta)$ for the path integral grand partition function $Z(\beta)$, yields the BCS-BEC gap and number equations
\begin{eqnarray}
\label{fixedpoints}
\gamma\left(\epsilon_F\right) \left(  \bar{\Psi}_1- \bar{\Psi}_2 \right) - \alpha\left(\epsilon_F\right)  \bar{\Psi}_1 +\beta\left(\epsilon_F\right) | \bar{\Psi}_1|^2  \bar{\Psi}_1 &=& 0, \nonumber \\
2\lambda_\nu\left(\epsilon_F\right)  \bar{\Psi}_2 +2 \chi\left(\epsilon_F\right)| \bar{\Psi}_2|^2 \bar{\Psi}_2 - \kappa\left(\epsilon_F\right)\gamma\left(\epsilon_F\right) \left( \bar{\Psi}_1- \bar{\Psi}_2 \right) &=& 0, \nonumber \\
2 \sigma^2(\epsilon_F) | \bar{\Psi}_2|^2 + \xi^2(\epsilon_F) | \bar{\Psi}_1|^2 -1 &=& 0.
\end{eqnarray}
Here, the new constant $\xi^2 \equiv \mu^2 {\partial_\mu a}$, $\epsilon_F=|\mu_F|$, and the overbars denote the equilibrium values of the order parameters. These equations can be solved numerically to yield the initial conditions for the order parameters. Impulse quenches of the detuning $\nu$ yield long term dynamics that can be approximated by linear displacements of $\Psi_{1,2}$ about radially fixed trajectories obtained from Eqs.~\ref{fixedpoints} \textit{viz.}
\begin{eqnarray}
 \Psi_1(t) &=& \left[r_\nu + \delta\Psi_1(t)\right]e^{-i\omega_\nu t},\nonumber \\
 \Psi_2(t) &=& \left[\left(1-\eta_\nu\right)r_\nu + \delta\Psi_1(t)\right]e^{-i\omega_\nu t},
\end{eqnarray}
where $\eta_\nu=\frac{\lambda_\nu/\kappa\gamma}{1+\lambda_\nu/\kappa\gamma}$, $r_\nu=\left[\frac{1}{\beta}(\alpha-\gamma\eta_\nu)\right]^{1/2}$ is the equilibrium value $|\bar{\Psi}_1|$ obtained from Eqs.~\ref{fixedpoints}, and the composite bosons are presumed to be noninteracting ($u_B=0$) for the sake of simplicity. Substituting this into the dynamics in Eqns.~\ref{dynamical_system}, and retaining only terms up to linear order in  $\mathcal{O}(|\delta\Psi_j|^2)$,  yields a system of  linear differential equations of the type $\delta \dot{x}_j \sim \sum_k \mathcal{J}^{jk}_\nu\delta x_k$. Here, $\delta\Psi_{1,2}=\delta x_{1,3} + i \delta x_{2,4}$, and ${j,k}=1-4$. The Jacobian matrix elements that describe this dynamics are given by $\mathcal{J}^{jk}_\nu\equiv\partial\dot{x}_j/\partial x_k$ at the fixed point. The linear dynamics is given by~\cite{mypaper4,strogatz:book} $|\delta x (t) \rangle \sim \sum^{4}_{j=1} c_j |\Omega^j_\nu\rangle e^{\Omega^j_\nu t}$, where $c_j$ are constants, $|\Omega^j_\nu\
rangle$ are the eigenvectors of $\mathcal{J}_\nu$  that correspond to eigenvalues $\Omega^j_\nu$, and $|\delta x (t) \rangle$ is the vector given by $\delta x_{1-4}$.
\begin{figure}
\resizebox{0.9\columnwidth}{!}{\includegraphics{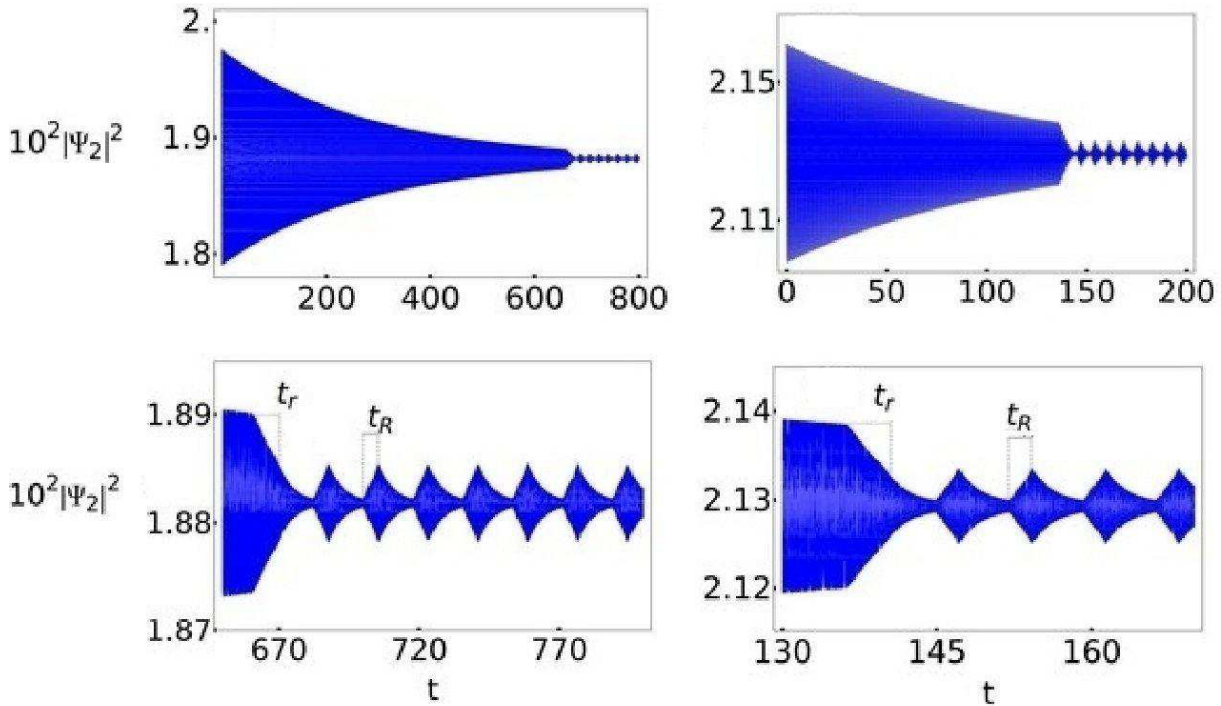} }
\caption{(Color online) Plots of the signal $|\Psi_2|^2(t)$  from the solutions to Eqs.~\ref{dynamical_system} for $u_B = 0$ after the system is quenched from a pure BEC to the shallow-BEC regime. Here, $\hbar = m =\mathcal{V} = 1$, and $u_F = -0.3$. The left panels
show plots of the time evolution for $g_r=25.0$ and $\nu = -55.0$, and the right panels show plots for $g_r=40.0$ and $\nu = -140.0$. The top panels show the time evolution for long times, and the bottom panels show the time evolution only after the onset of dynamics of $\mathcal{O}(|\delta\Psi_2|)$. The presence of partial collapse and revival of the matter wave can be noted after a relaxation time $t_r$ as indicated in the bottom panels. The revival times $t_R$ are also indicated in these figures. In the left panels, $t_r\simeq9$, $t_R\simeq6$, and in the right panels, $t_r\simeq4$, $t_R\simeq2.5$. Reproduced from~\cite{mypaper4} with permission of author.}
\label{fig:colrev}
\end{figure}
The eigenvalues $\Omega_\nu$ are given by the characteristic equation $|\mathcal{J}_\nu-\Omega_\nu I|=0$. In this case, the equation is biquadratic of the type $\Omega_\nu^4 + \mathcal{B}_\nu\Omega_\nu^2-\mathcal{C}_\nu =0$, where the constants $ \mathcal{B}_\nu$ and $ \mathcal{C}_\nu$ are functions of $\alpha,\beta,\gamma,\kappa$, and $\nu$~\cite{mypaper4}. The primary quantity of interest is the \textit{discriminant} of the quartic equation $\mathcal{D}_\nu=1+\left(4\mathcal{C}_\nu/\mathcal{B}^2_\nu\right)$. The roots of $\Omega_\nu$ can be written in terms of this discriminant as $\Omega_\nu=\pm\sqrt{-\mathcal{B}_\nu/2}\left(1\pm\sqrt{\mathcal{D}_\nu}\right)^{1/2}$. The sign of  $\mathcal{D}_\nu$ determines whether the roots are real or complex. Thus, it plays a key role in determining the nature of these trajectories. If $\Omega_\nu$ are pure imaginary, a 'deep-BEC' regime can be identified where the trajectories assume orbital motion about the radial fixed points. The orbits constitute two oscillations 
of frequencies $\Omega^{\pm}_\nu=|\sqrt{-{\mathcal{B}_\nu}/{2}}(1\pm\sqrt{\mathcal{D}_\nu})^{1/2}|$. Their combination results in periodicity iff $\Omega^\pm_\nu$ have rational winding numbers $n^+/n^- = \Omega^+_\nu/\Omega^-_\nu$, where $n^\pm$ are integers with no common factors. The number density of the BEC, $|\delta \Psi_2|^2=\delta x^2_3 + \delta x^2_4$, will have $4$ oscillations, $2\Omega^\pm_\nu$ and $\Omega^+_\nu\pm\Omega^-_\nu$. When the $\Omega_\nu$ start to pick up real parts, then a 'shallow BEC' regime can be identified where the trajectories corresponding to negative real parts are stable (the eigenvalues with positive real parts correspond to displacements that violate number conservation and are unphysical) and decay in spirals to the orbits of the radial fixed points, thus creating a stable limit cycle. Recently, Hopf bifurcations that are similar to those discussed above have been studied in the more general dynamics of Eqs.~\ref{timmermans:fulldyn}~\cite{hopf}.

A rapid quench from the 'deep BEC' regime to the 'shallow BEC' regime can thus be expected to yield orbital phase space dynamics of $\Psi_j$ around the equilibrium fixed points after a sufficiently long time, producing a 'collapse and revival' of the matter wave packet. The dynamics has been analyzed numerically for an ultracold gas of $100$ fermions in a box of unit volume for larger values of $\epsilon_F$ in the shallow-BEC regime by Roy~\cite{mypaper4} and the ensuing dynamics of a representative parameter set are plotted in Fig.~\ref{fig:colrev}. For long times, note from the panels in this figure that a sudden onset of partial \textit{collapse and revival} of the matter wave begins after a short initial relaxation time $t_r$. The onset of this collapse and revival takes place at around $10^3$ numerical units. Also, note from these panels that all three time scales, $t_d$, $t_r$ and $t_R$, are larger for smaller $g_r$. Thus, it is expected that $t_d \rightarrow \infty$ as $g_r \rightarrow 0$, removing 
the collapse and revival effect in the single channel case. 
\section{Microscopic Dynamics }
\label{sec:fermidyn}

 {This section details studies of the full microscopic dynamics of responses in a single channel fermionic superfluid in the BCS regime with time dependent parameters. The dynamics of the superconducting BCS state in metals has been of interest for a long time}~\cite{tinkham},  {in particular, the regime where the BCS quasiparticles relax slowly compared to the gap} (Anderson~\cite{anderson}).  {The regime of interest in this section is one where the external parameters can be changed rapidly compared to the equilibrium BCS relaxation time ($\hbar\Delta^{-1}$) but slowly compared to the relaxation of the quasiparticle excitations. This allows for the analytic continuation in time of the equilibrium BCS problem. The resultant dynamical system consists of the Schr\"odinger equations of the microscopic wavefunction, and a time-dependent BCS gap equation that is self-consistently coupled to the Schr\"odinger equations. Among the time dependencies of interest are variation of the four fermion interaction 
amplitude via the Feshbach detuning, as well as variation of the chemical potential. Both of these are straightforward to perform in experiments. The four-fermion amplitude can be rendered time dependent by varying the Feshbach field in time adiabatically (relative to the scattering continuum), and the chemical potential can be effectively varied in time by confining the system in a large isotropic trap with a time-dependent intensity, stiffness, or minima}~\cite{colrev} {. The detuning of the confining lasers from the internal degrees of freedom of the confined atoms can also be varied in time to produce effectively time dependent chemical potentials}~\cite{fdet}.

 {In BCS systems modeled by self-consistent dynamics, both sudden and periodic quenches can be modeled by the BCS Hamiltonian}
\begin{equation}
H(t) = \sum_{{\bf k}\sigma} \left[ f_{\bf
k}-\mu(t)\right]c^\dagger_{{\bf k}\sigma}c_{{\bf k}\sigma}-g(t) \sum_{{\bf k}, {\bf k'},{\bf k''}} c^\dagger_{{\bf k}+{\bf
k''}\uparrow} c^\dagger_{{\bf k'}-{\bf k''}\downarrow} c_{{\bf
k'}\downarrow} c_{{\bf k}\uparrow}. \label{fermiham}
\end{equation}
 {Here, $f_{\bf k}=\epsilon_{\bf k}-\mu_0$ are the band energies relative to equilibrium chemical potential $\mu_0$, and $c_{{\bf k}\sigma}$ represent the annihilation operators for fermions of momentum ${\bf k}$ and spin $\sigma=\{\uparrow, \downarrow\}$. The first term represents the kinetic energy of the fermions, and the second term the four-fermion interaction energy with amplitude $g=g(0)>0$ which represents attractive interaction between the fermions. In the rest of this work, the trap potential is assumed to vary sufficiently slowly that a locally constant chemical potential $\mu_0 = \epsilon_{k_F}$ can be used to describe the fermions in the trap. At $t=0$, this system can be mapped from the BCS Hamiltonian via  a Hubbard-Stratonovich transformation}~\cite{hubbard:hstrans,coleman}  {to yield a BCS-superfluid state from the Bogoliubov de-Gennes equations,}
\begin{eqnarray}
E({\bf k})\left(\begin{array}{c} u_{\bf k} \\ v_{\bf k}
\end{array} \right) &=& \left(
\begin{array}{cc}
(\epsilon_{\bf k}-\mu_0) & \Delta({\bf k})\\
\Delta^{\ast}({\bf k}) & -(\epsilon_{\bf k}-\mu_0)
\end{array} \right) \left(\begin{array}{c} u_{\bf k} \\ v_{\bf k}
\end{array} \right), \nonumber\\ \label{bdg1}
\end{eqnarray}
where $u_{\bf k}$ and $v_{\bf k}$ are the amplitudes of the particle and the hole in a BdG quasiparticle and are related to the BCS
wavefunction by
\begin{eqnarray}
|\psi\rangle &=& \prod_{\bf k} (u_{\bf k} + v_{\bf k} c_{\bf
k}^{\dagger} c_{{\bf -k}}^{\dagger}) |0\rangle. \label{bcswav}
\end{eqnarray}
The pair-potential $\Delta({\bf k})$ depends on the pairing symmetry and is given by
\begin{eqnarray}
\Delta({\bf k}) &=& \Delta_0, \quad {\rm s-wave }, \nonumber\\
\Delta({\bf k}) &=& \Delta_0 [\cos(k_x) -\cos(k_y)],  \quad {\rm
d}_{x^2-y^2}{\rm -wave}. \label{pp}
\end{eqnarray}
For s-wave pairing, the pair potential satisfies the self-consistency relation
\begin{eqnarray}
\Delta_0 &=& g \sum_{\bf k} u_{\bf k}^{\ast} v_{\bf k}.
\label{self1}
\end{eqnarray}
Eq.~\ref{bdg1} and~\ref{self1} admit the well-known BCS solution
\begin{eqnarray}
E({\bf k})&=& \pm \sqrt{(\epsilon_{{\bf k}}-\mu_0)^2+|\Delta_0|^2}
\nonumber\\
u^{\rm eq}_{\bf k} (v^{\rm eq}_{\bf k}) &=& \frac{1}{{\sqrt 2}}
\left( 1+(-) \frac{(\epsilon_{\bf k} -\mu_0)}{E({\bf k})}
\right)^{1/2}. \label{eqsol}
\end{eqnarray}
As the system evolves in time, it can be modeled by the time dependent Bogoliubov de-Gennes equations \textit{viz.}
\begin{eqnarray}
i \partial_t \left(\begin{array}{c} u_{\bf k}(t) \\ v_{\bf
k}(t)
\end{array} \right) &=& \left(
\begin{array}{cc}
[f_{\bf k}-\mu (t)] & \Delta({\bf k;t})\\
\Delta^{\ast}({\bf k};t) & -[f_{\bf k}-\mu(t)]
\end{array} \right) \left(\begin{array}{c} u_{\bf k}(t)\\ v_{\bf
k}(t)
\end{array} \right), \label{bdg2} \nonumber\\
\end{eqnarray}
together with the self-consistency condition, which reads
\begin{eqnarray}
\Delta({\bf k};t) &\equiv& \Delta(t) = g(t) \sum_{\bf p} u^{\ast}_{\bf
p}(t) v_{\bf p}(t),\label{tdsc1}
\end{eqnarray}
for s-wave pairing. Here, $\hbar=1$, and $u_{\bf k}$ and $v_{\bf k}$ are the amplitudes of the particle
and the hole in a BdG quasiparticle.  { As in the case described in section~\ref{sec:myoldstuff}, detailed justifications of the time dependent Bogoliubov de-Gennes equations can be obtained from the nonequilibrium Schwinger-Keldysh formalism by ignoring the collision integrals in the Dyson equation and building the particle-hole amplitudes from the Keldysh Greens function after applying the approximations described in the first paragraph of this section. See, for instance, refs}~\cite{gorkov} and ~\cite{volkov} (Eqs. $1$-$20$)\footnote{Note that performing the substitutions $f\rightarrow u^\ast_{\bf k} v_{\bf k}$, $g\rightarrow 1-2|v_{\bf k}|^2$, $\xi\rightarrow f_{\bf k}-\mu$, and $\int \mathrm{d}\xi\rightarrow\sum_{\bf k}$ in Eqs. $20$ of ref~\cite{volkov} yield the BdG dynamics in Eqs.~\ref{bdg2} and~\ref{tdsc1} above.}

 {The dynamics of the gap parameter in such regimes have been investigated by Szymanska \textit{et. al.}}~\cite{szymanska},  {Barankov \textit{et. al.}}~\cite{barankov}, {Andreev \textit{et. al.}}~\cite{andreev:noneqmbcsbec}, {Pekker \textit{et. al.}}~\cite{pekker},  {using various trial solutions. The ensuing dynamics has, in particular, been cited as a way to identify pairing symmetries in Fermi superfluids}~\cite{pekker,ourpaper} {, as well as Stoner instabilities}~\cite{stoner} {, and investigations of quench dynamics across the BCS-BEC crossover}~\cite{stoner} {, where a renormalized $g(t)$ switches sign for $T=0$ in the adiabatic limit}~\cite{bcsbec}.  { The method  used by Barankov, Levitov and Spivak (as well as a similar approach by Andreev, Gurarie and Radzihovsky}~\cite{andreev:noneqmbcsbec}) {can be generalized in order to obtain the dynamics for $\mu(t)=0$, $g(t)=g_0\theta(t)$}~\cite{barankov} {. The unpaired state of no Cooper pairs, given by  $u_{\bf k}(t)=\theta(f_{\bf k}) 
e^{-if_{\bf k}t}$, $v_{\bf k}(t)=\theta(
-f_{\bf k}) e^{if_{\bf k}t}$, satisfies Eqs.}
~\ref{bdg2}  {with $\Delta(t)=0$ $\forall t$ from Eq.}~\ref{tdsc1}.
 {Thus, this trajectory acts as a generalized fixed point of the dynamics in the form of a limit cycle. A linear stability analysis around this cycle was analyzed by Abrahams and Tsuneto}~\cite{abrahams} {, revealing that $|\Delta(t)|$ grows as $e^{\Delta_0 t}$ for linear order displacements about the solutions of $u_{\bf k}(t)$ and $v_{\bf k}(t)$ above. Defining the variable
$w_{\bf k}=(u_{\bf k}/2v_{\bf k})(1+{\rm sign}f_{\bf k})+(v_{\bf k}/2u_{\bf k})(1-{\rm sign}f_{\bf k})$, the dynamics (for $f_{\bf k}>0$) can be written as}
\begin{equation}
\label{wdyn}
i \dot{w}_{\bf k}= 2\left[f_{\bf k}-\mu(t)\right] w_{\bf k} + \Delta(t) - \Delta^\ast(t)w_{\bf k},
\end{equation}
 {with a similar equation for  $f_{\bf k}<0$. Substituting $\Delta(t)=e^{i\Omega t}/\alpha(t)$ into Eq.}~\ref{wdyn} { and using
the fact that $w_{\bf k}=2[f_{\bf k}-\mu(t)]/\Delta^\ast-i\partial/\partial t (1/\Delta^\ast)$ yields}
\begin{equation}
\label{alphadyn}
\alpha\dot{\alpha}^2+\alpha^\ast+|\alpha|^2\left(2i\alpha\dot{\mu}-\ddot{\alpha}\right)=4\left(\alpha^2-|\alpha|^2\right)\left[f_{\bf k}-\mu(t)\right]\bigg\{\alpha\left[f_{\bf k}-\mu(t)\right]+i\dot{\alpha}\bigg\}.
\end{equation}
 {The ansatz $\alpha(t)\in\mathbb{R}$ $\forall t$, causes the RHS of Eq.}~\ref{alphadyn} { to vanish, yielding a nonlinear differential equation for the dynamics of $\alpha$ (and hence $\Delta$) that is independent of ${\bf k}$ and decoupled from Eqs.}~\ref{bdg2} {, $\alpha\ddot{\alpha}-\dot{\alpha}^2-1=2i\alpha^2\dot{\mu}$. In general, this leads to a contradiction, since this equation implies that $\alpha(t)$ picks up an imaginary part, and relaxing the assumption $\alpha(t) \in \mathbb{R}$ causes momentum dependent terms in the RHS of Eq.}~\ref{alphadyn}  {to remain. Thus, this treatment does not yield a closed form for the dynamics of $\Delta(t)$. However, if $\mu(t)=0$, Eq.}~\ref{alphadyn}  {can be readily integrated with the ansatz  $\alpha(t)\in\mathbb{R}$ $\forall t$}~\cite{barankov} { to yield $\Delta(t)=\Delta_0 e^{-i\Omega t}/\cosh{\Delta_0 t}$. The frequency $\Omega$ can be obtained from the initial condition $\dot{\Delta}(0)=\Delta_0 \dot{g}(0)/g$ (obtained by applying Eqs.}~\ref{eqsol}
 {  to Eqs.}~\ref{bdg2}  {and the time derivative of Eq.}~\ref{tdsc1}  {at $t=0$) to yield $\Omega=-i \dot{g}(0)/g$. Given that $g(t)$ must be real for the BCS Hamiltonian to be Hermitian, this leads to the solution}
\begin{eqnarray}
\label{delsoln}
\Delta(t)&=&\frac{\Delta_0 e^{-\Gamma t}}{\cosh{\Delta_0 t}},\nonumber\\
\Gamma   &=& {\frac{\partial}{\partial t}\ln{g(t)}\bigg|_{t=0}}.
\end{eqnarray}
 {If $\Gamma\geq0$, the gap parameter starts from $\Delta_0$ and damps out to $0$, stabilizing on the fixed point solution shown earlier in the para. However, if $\Gamma<0$, then, for large times $t\gg\Delta^{-1}_0$, $\Delta(t)\sim 2\Delta_0 e^{-(\Gamma+\Delta_0)t}\rightarrow 0$ if $|\Gamma|<\Delta_0$, $\Delta(t)\rightarrow 2\Delta_0$ if $|\Gamma|=\Delta_0$, and $\Delta(t)$ blows up if $|\Gamma|>\Delta_0$. The first of the three cases can be readily solved in the asymptotic limit. In the second case, Eqs.}~\ref{bdg2}  {decouple from each other at large times with $\Delta(t)\approx 2\Delta_0$, and become \textit{non self-consistent}  linear Schr\"odinger equations with twice the equilibrium gap. The latter-most case is clearly unphysical due to norm conservation , $|u_{\bf k}(t)|^2+|v_{\bf k}(t)|^2=1$ $\forall {\bf k}$ and $t$. 
Investigations of this regime are a subject of further study.}

 {In the other two cases, the solution in Eq.}~\ref{delsoln}  {can be used to integrate Eqs.}~\ref{wdyn} {. This yields $w_{\bf k}(t)=e^{-ia_{\bf k}(t)}\left[d_{\bf k}-i\int\mathrm{d} t\Delta(t)e^{ia_{\bf k}(t)}\right]$ for $f_{\bf k}>0$, with a similar solution for $f_{\bf k}\leq0$. Here, $d_{\bf k}$ is a constant determined from the initial conditions, and $a_{\bf k}(t)=\int \mathrm{d} t \left[f_{\bf k}-\Delta(t)\right]$ can be evaluated in terms of Hyper-geometric polynomials. Analytical solutions to these equations have been provided by Barankov \textit{et. al.} using Bloch vectors}~\cite{barankov} {. The special case of a rapid quench in $\Delta$ deserves special interest. If the time that the gap takes to relax from $\Delta_0$ to the final value $\Delta_1$ 
(where $\Delta_1=0$ if $\Gamma\geq0$ or $-\Delta_0<\Gamma<0$, and $\Delta_1=2\Delta_0$ if $\Gamma=-\Delta_0$) is neglected, then the final non self-consistent Schr\"odinger equations are those of the time independent Hamiltonian $H^{(1)}_{\bf k}={\bf T}_{\bf k}\cdot{\bf E}_1({\bf k})$, where ${\bf T}_{\bf k}=\tau^x_{\bf k}\hat{x}+\tau^y_{\bf k}\hat{y}+\tau^z_{\bf k}\hat{z}$, with $\tau^{x,y,z}_{\bf k}$ being the three Pauli matrices in particle-hole space, and the vector  ${\bf E}_1({\bf k})=f_{\bf k}\hat{z}+\Delta_1\hat{x}$ with magnitude $E_1({\bf k})=\sqrt{f^2_{\bf k}+\Delta^2_1}$. The initial state $|\psi_{\bf k}(0)\rangle$ can be computed from the initial conditions for $u^{\rm eq}_{\bf k}$ and $v^{\rm eq}_{\bf k}$ given by Eqs.}~\ref{eqsol} {. Thus, the solution to the particle-hole amplitudes in this limit is
$|\psi_{\bf k}(t)\rangle=\exp{\left(iH^{(1)} t\right)}$ $|\psi_{\bf k}(0)\rangle$. This can be expanded using $SU(2)$ algebra, and can be used to determine the particle and hole amplitudes.}
 {If $\Delta_1=0$, then the system merely undergoes phase oscillations and the particle-hole amplitudes  are frozen at their equilibrium values. If $\Delta_1=2\Delta_0$, the particle-hole amplitudes can be averaged out over long times}
 {by averaging out the harmonic terms in their exact dynamics.} 
 {Substituting the values of the equilibrium amplitudes from Eq.}~\ref{eqsol}  {yields the long time average(s) to be $|u^s_{\bf k}|^2 (|v^s_{\bf k}|^2)$, where}
\begin{equation}
u^s_{\bf k} (v^s_{\bf k})=\frac{1}{\sqrt{2}}\left(1+(-)\frac{f_{\bf k}}{\sqrt{f^2_{\bf k}+\Delta^2_0}}\quad\frac{f^2_{\bf k}+2\Delta^2_0}{f^2_{\bf k}+4\Delta^2_0}\right)^{1/2}\approx \frac{1}{\sqrt{2}}\left(1+(-)\frac{f_{\bf k}}{2\sqrt{f^2_{\bf k}+\Delta^2_0}}\right)^{1/2}.
\end{equation}
 {Here, the approximation holds for large gap $\Delta_0$. Thus, the system appears to attain a BCS-like steady state given by $|\psi^s\rangle = \prod_{\bf k} (u^s_{\bf k} + v^s_{\bf k} c_{\bf k}^{\dagger} c_{{\bf -k}}^{\dagger}) |0\rangle$, up to a phase. If $\Delta_0$ is sufficiently large, then this state does not differ much from the equilibrium state as can be seen from the equation above. The steady state magnetization, defined by $Q=1-2\sum_{\bf k}|v^{s}_{\bf k}|^2$, is clearly about half of the equilibrium magnetization $Q^{\rm eq}=1-2\sum_{\bf k}|v^{\rm eq}_{\bf k}|^2$ for a sufficiently large gap. Thus, equilibrium states with half-filled symmetry retain this symmetry in the long time dynamics of the magnetization.

The dynamics of impulse quenches in $g$ can be formulated as a special case of the para above. Here, $\Gamma=0$, and $g$ is a constant for $t>0+$. The system is populated in a BCS state characterized by a gap $\Delta=\Delta_0$. Barankov \textit{et. al.}}~\cite{barankov} { analyzed the ensuing soliton dynamics for various values of $\Delta_0$ from Eq.}~\ref{delsoln} { based on the linear stability analysis detailed above. This yields elliptic oscillations that cause overlapping solitons in $u_{\bf k},v_{\bf k}$, causing them to oscillate about those for an ideal Fermi gas. Volkov and Kogan}~\cite{volkov} {, have obtained generalized fixed points of}~\ref{bdg2} { (with nonzero $\Delta_0$), of the type $u^\ast_{\bf k}v_{\bf k}\sim -i\Delta_0 \phi_{\bf k}$, $1-2|v_{\bf k}|^2\sim -if_{\bf k}\phi_{\bf k}$ with real amplitude $\phi_{\bf k}$, that restrict the Hilbert space to surfaces of constant $u^\ast_{\bf k}v_{\bf k}$, $v^\ast_{\bf k}v_{\bf k}$. A linear stability analysis of Eqs.}~\ref{bdg2}  {around 
this 
trajectory revealed bifurcations in the dynamics of $\Delta(t)$ that cause its trajectory to change from slow relaxations to oscillatory decays of rate  $\mathcal{O}(t^{-1/2})$ towards $\Delta_0\neq 0$. The study of general time dependent $g$ under these conditions is a subject of future study.}

The dynamics of $u_{\bf k}$ and $v_{\bf k}$ in the special case when $g(t)$ is fixed at $g_0=g$, and  $\mu(t)=\mu_a\sin{\omega t}$ is the time periodic drive supplied to the system, has acquired renewed interest, and will be discussed below.  {In this regime, the analysis of Barankov \textit{et. al.} detailed in previous paragraphs breaks down, necessitating other treatments for the ensuing dynamics. Such dynamics have been investigated in the context of Cooper-pair scattering by Challis \textit{et. al.}}~\cite{challis}  {using Bloch-Fourier series expansions for $u_{\bf k}, v_{\bf k}$ and $\Delta(t)$, and truncating the series terms, obtaining approximate solutions that demonstrate Bragg scattering of correlated atom pairs. Studies of soliton formation in such systems at equilibrium due to the occurrence of Andreev reflection  have also been performed by Antezza \textit{et. al.}}~\cite{antezza} {, and the traveling dynamics of such solitons of the type $\Delta(x,t)=\Delta(x-vt)$ for different pairing 
symmetries have been 
reported by Scott \textit{et. al.}}~\cite{scott} {, Liao and Brand}~\cite{liao} {. 
Roy \textit{et. al.}}~\cite{ourpaper}  {have computed the response of the system as it evolves in time, demonstrating  that the BCS self-consistency condition introduces nonlinearities in the dynamics and shapes the long-time behavior of the fermions subjected to the drive.}
\begin{figure}
\resizebox{0.5\columnwidth}{!}{\includegraphics{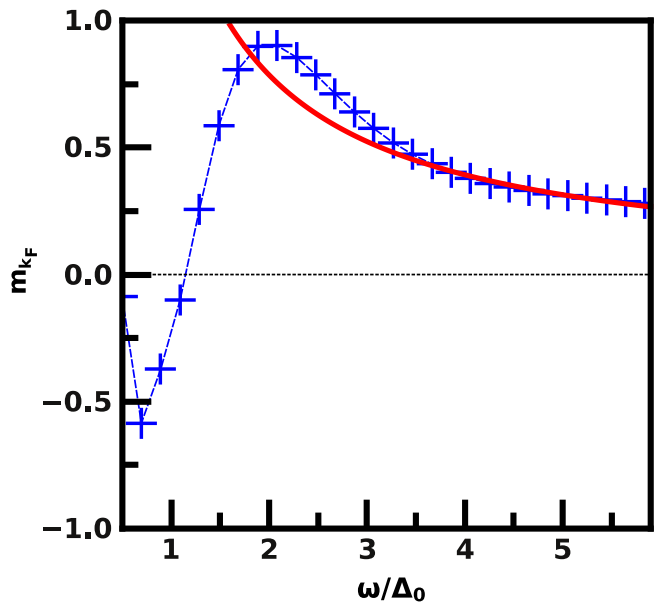} }
\caption{(Color Online) Numerical plots (blue and crossed) of Fermi surface magnetization after a double passage of the avoided crossing as a function of the drive frequency. The system parameters are $\Delta_0=\mu_a=0.1$, $\mu_0=0.01$, and the lattice size is $N=144\times 144$ in two dimensions. This is contrasted with a plot of the analytical formula of $m_{k_F}$ (red) obtained in section~\ref{sec:fermidyn}. Reproduced from~\cite{ourpaper} with permission from the authors.}
\label{fig:magcomp}
\end{figure}
\begin{figure}
\resizebox{0.75\columnwidth}{!}{\includegraphics{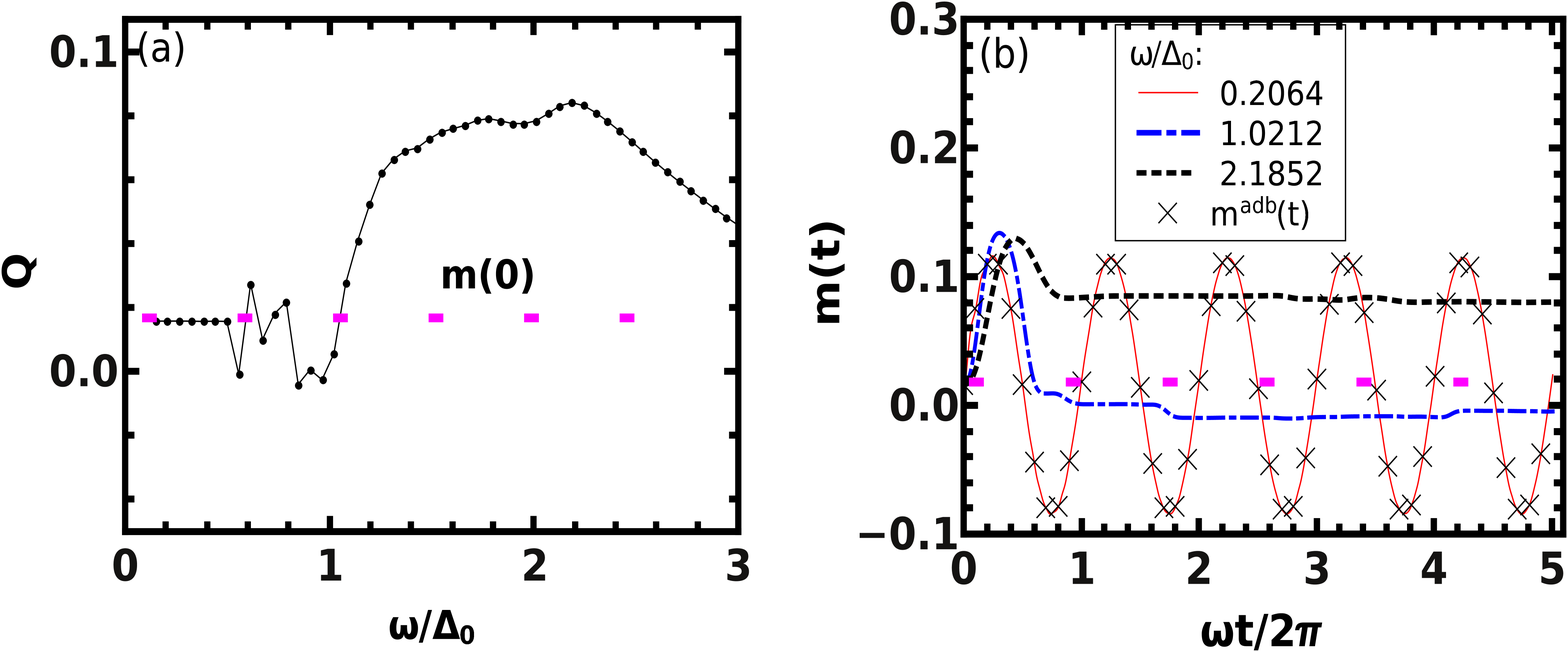} }
\caption{(Color online) Left Panel: Plot of Q, the time average of magnetization, as a function of $\omega$, with averaging carried over $10$ drive cycles. Right panel: Plot of $m(t)$ as a function of $\omega t/2\pi$ for representative values of $\omega$ indicated in the legend. A few representative values of the adiabatic magnetization $m^{adb}(t)$ is shown using crosses. In both panels, the equilibrium magnetization $m(0)$ is indicated by a solid (red) horizontal line and all parameters are same as in Fig.~\ref{fig:magcomp}. Reproduced from~\cite{ourpaper} with permission from the authors.} 
\label{fig:mags}
\resizebox{0.75\columnwidth}{!}{\includegraphics{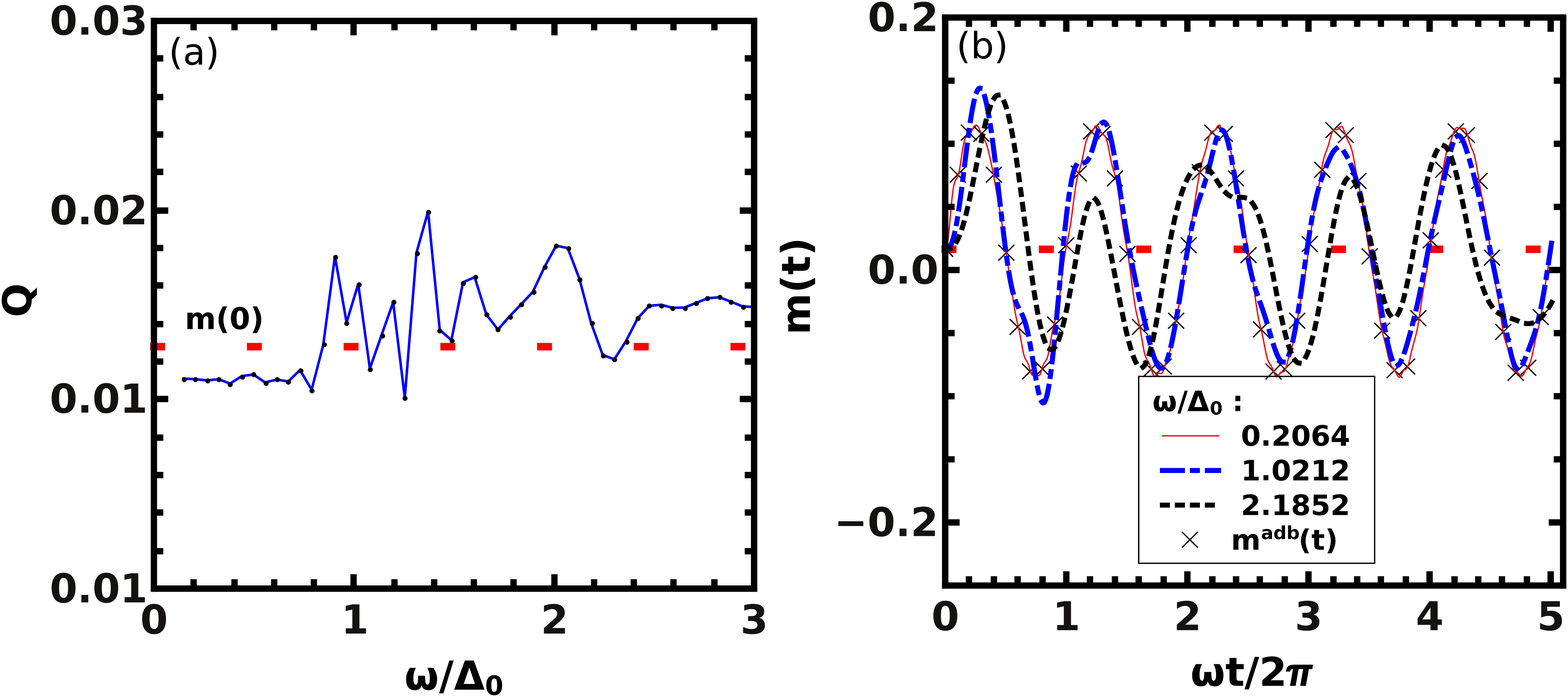} }
\caption{(Color online) Same as in Fig.~\ref{fig:mags} but for the
non-self-consistent dynamics. Reproduced from~\cite{ourpaper} with permission from the authors.} \label{fig:mag:nsc}
\end{figure}
 {If the system is started from the state in Eq.}~\ref{eqsol}  {at $t=0$ and evolved in time, a perturbative analysis of the ensuing dynamics is possible. The remaining part of this section focuses on the salient features of such a treatment done by Roy \textit{et. al.}} ~\cite{ourpaper} {. The possibility of performing a linear stability analysis around a texture state similar to that by Volkov and Kogan as mentioned in the previous paragraph is a subject of future study.}

The primary response of interest is the effective 'magnetization'~\cite{arnab1}, defined as $m(t)=\sum_{\bf k}m_{\bf k}(t)$, where $m_{\bf k}(t)=\langle \tau^z_{\bf k} \rangle(t)$. The self-consistent dynamics given by Eqs.~\ref{bdg2} and~\ref{tdsc1} can be perturbatively integrated, using the Zener approximation~\cite{wittig:lzformula}, for one passage across the avoided crossing in the adiabatic spectrum. The adiabatic state, is given by the equilibrium state in Eq.~\ref{eqsol} continued in time via the transformation
$\mu_0\rightarrow\mu_0+\mu_a\sin{\omega t}$, and the adiabatic spectrum by the energies $\pm E({\bf k})$ transformed similarly. This integration has been done in a manner similar to~\cite{wittig:lzformula} except with a time dependent gap $\Delta(t)$. This yields a fermion density (which goes as $|v^{(1)}_{\bf k}|^2$) at the Fermi surface (given by $f_{\bf k}=0$) after a single passage to be
\begin{equation}
\label{eq:lz:pert}
|{v}^{(1)}_{\bf k}|^2  =  \frac{1}{2}\times\bigg\{1-2\chi_0 |c_{f_{\bf k}}|^2+2\sqrt{\chi_0}\times{\rm Re}\left[\left(1+i\right)c_{f_{\bf k}}\right]\bigg\}, 
\end{equation}
where
\begin{equation}
\label{eq:ck0}
c_{f_{\bf k}} \equiv \sum^\infty_{n=0} \frac{1}{n!}\times\frac{\left(-i\right)^n}{\left(4\mu_a\omega\right)^n} \times \frac{\partial^{2n}\theta_{\bf k}}{\partial^{2n}t}\bigg|_{t=0}.
\end{equation}
Here $\chi_0=\pi\Delta^2_0/(\mu_a\omega)$, and $\theta_{\bf k}(t)=v_{\bf k}(t)\times[\Delta(t)/\Delta_0]\times\exp{\{i\int^t_0\mathrm{d}t' \mu(t')\}}$. For the non self-consistent case, $\Delta(t)=\Delta_0$ $\forall t$, and the series in Eq.~\ref{eq:ck0} yields the traditional Landau Zener formula~\cite{wittig:lzformula}, $|v^{(1)}_{\bf k}|^2\sim e^{-\chi_0}/2$. For a self-consistent gap given by Eq.~\ref{tdsc1}, 
the quantities $c_{\bf k}$ and $|{v}^{(1)}_{\bf k}|^2$ can be approximated by truncating Eqs.~\ref{eq:ck0} and~\ref{eq:lz:pert} to lowest order in $\chi_0$, yielding a fermion density that differs from that of the non self-consistent case. The density at the Fermi surface after two complete passages (amounting to one period of the drive) across the avoided crossing can be approximated by~\cite{review:lzstls} $ |{v}^{(2)}_{\bf k}|^2 = 2 |{v}^{(1)}_{\bf k}|^2 \left(1-|{v}^{(1)}_{\bf k}|^2\right)$,
where St\"uckelberg interferometry has been neglected for large $\omega$. The resultant density, written in terms of the 'effective magnetization' $m_{k_F} \equiv \langle\tau^z_{{\bf k}=k_F}\rangle  = 1-2|{v}^{(2)}_{\bf k}|^2$, approximately goes as ${\chi_0}/{2}$. This yields a scaling law for this magnetization $m_{k_F}\sim\omega^{-1}$, that is similar to that of the non self-consistent case (as can be seen by Taylor-expanding the Landau Zener formula), but with a comparatively smaller scaling constant (by a factor of $4$ to the lowest order). This result has been compared to numerical simulations in Fig.~\ref{fig:magcomp}. The two diverge for $\omega\lesssim\Delta_0$ since that regime is nonperturbative, but agree quite well for larger $\omega$.

Numerical simulations of the dynamics above have been performed by Roy \textit{et. al.}~\cite{ourpaper} for longer times while including the entire Brillouin zone in the calculation of the response. Plots of the instantaneous magnetization and its time average (over long times) in Fig.~\ref{fig:mags} show many differences with the non self-consistent case, where Landau - Zener oscillations~\cite{review:lzstls} and other dynamical phenomena such as 'exotic freezing' of the response~\cite{arnab1} are expected. The primary difference is the appearance of a 'steady state' in the magnetization for the self consistent case when $\omega\gtrsim\Delta_0$. Smaller $\omega$'s yield results similar to the non self-consistent case, especially when $\omega\ll\Delta_0$. In that regime, the magnetization approaches that of the adiabatic state ( which is defined in the fourth paragraph). The adiabatic magnetization, $m^{adb}$, is shown as black crosses in the right panel of Fig.~\ref{fig:mags}. This is consistent with the 
quantum adiabatic theorem, where adiabatic dynamics is expected if the 
time-scale of the dynamics ($2\pi/\omega$) is greater than the relaxation time $\Delta^{-1}_0$. 

The above behavior is to be contrasted with the {\it non-self-consistent} case summarized in Fig.~\ref{fig:mag:nsc}.  {In the non self-consistent case, Eq.}~\ref{self1}  {is not analytically continued in time and the gap $\Delta(t)$ is kept constant at $\Delta_0$. In this case, the dynamics is described solely by Eqs.}~\ref{bdg2}  {without the self-consistent update in Eq.}~\ref{tdsc1} {. Such dynamics can be described by the $1$-D BCS reduced model}~\cite{agarg} {, whose co-ordinate space Hamiltonian}
\begin{equation}
H(t)=-\sum_i \left[c^\dagger_ic_{i+1} + \Gamma c^\dagger_ic^\dagger_{i+1}\right]+\frac{\mu(t)}{2} \sum_i \left(1-2 c^\dagger_ic_i\right),
\end{equation}
 { can be connected to the anisotropic quantum XY model. A Jordan Wigner transformation of this Hamiltonian at $t=0$}~\cite{katsura} {, yields an $SU(2)$ Hamiltonian similar to that in Eq.}~\ref{bdg1} { with $\epsilon_k=-(\mu_0/2)+\cos{k}$ and $\Delta(k)=\Gamma\sin{k}$, where $\mu_0=\mu(0)$. The dynamics can thus be described by Eq.}~\ref{bdg2}  { sans the self consistency in Eq.}~\ref{tdsc1} {. This causes the system to decouple into $\mathcal{N}$ independent two-level systems (henceforth abbreviated as TLS), each characterized by momentum $k$. The study of the periodic dynamics of quantum XY systems using TLS dynamics was performed by Mukherjee and Dutta~\cite{xyseminal} . In addition, such systems can also be generated from Ising} (in $1$-D, see  
Dziarmaga~\cite{dziarmaga}, Cherng and Levitov~\cite{cherng:levitov}, and Sen \textit{et. al.},~\cite{dsen})  {or Kitaev} (in 2-D, see Chen and Nussinov~\cite{kitaev2d}, Sengupta \textit{et. al.}~\cite{kitaev2d:krishda}, Mondal \textit{et. al.}~\cite{sreyoshi})   {models via the Jordan Wigner transformation, and their dynamics described by those of quantum TLS systems.}

Numerical solutions to such TLS dynamics for the isotropic case ($\Delta({\bf k})=\Delta_0$) in $2$-D are shown in Fig.~\ref{fig:mag:nsc} for the same parameters as those used for the self-consistent dynamics earlier in this section. Here, $m(t)$ always executes a large, almost synchronized oscillation, approximately following the adiabatic path (black crosses in the right panel of Fig.~\ref{fig:mag:nsc}). Naturally, the resulting values of $Q$ are close to $m(0)$ (albeit with some small fluctuations). This suggests that the synchronous oscillation is simply a manifestation of the near-adiabatic nature of the dynamics. The qualitative departure from this behavior in the self-consistent case seems to stem form the non-adiabaticity induced by the self-consistency condition (Eq.~\ref{tdsc1}) which makes the effective Hamiltonian non-linear. The overlapping eigenfunctions of the non-linear Hamiltonian makes the criteria for adiabatic behavior much more restricted compared to a linear case (see. e.g., Yukalov~\
cite{Yukalov}).

 {The quantum dynamics of driven TLS systems of the type described in Eq.}~\ref{bdg1}  {can be solved analytically using various techniques.
The BCS solution in Eq.}~\ref{eqsol},  {when continued in time via $\mu_0\rightarrow\mu(t)$, describe the solutions to Eq.}~\ref{bdg1} { in the adiabatic limit, when the characteristic time scale of the variation of $\mu(t)$ ($\omega^{-1}$ if $\mu(t)=\mu_a\sin{\omega t}$) is much larger than $\Delta^{-1}$.}
\begin{figure}
\resizebox{0.5\columnwidth}{!}{\includegraphics{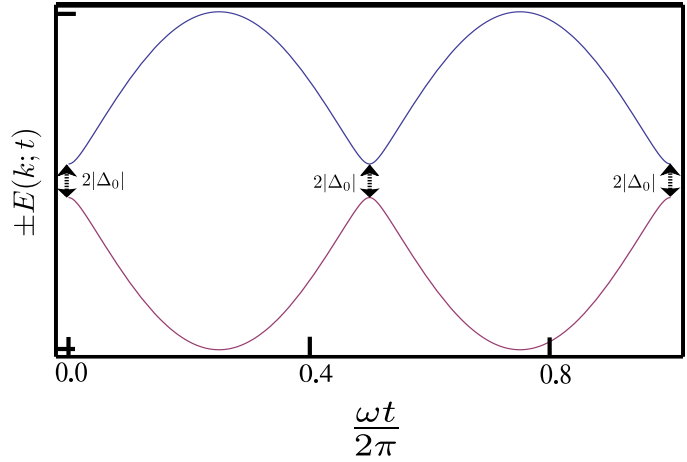} }
\caption{(Color online) Plots of the adiabatic energy levels $\pm E(k;t)$ as functions of dimensionless time illustrating the adiabatic impulse model in Landau-Zener theory in the non self-consistent case. The plots are for $f_{\bf k}=0$, $\Delta({\bf k})=\Delta_0=\mu_a/10$, with $\mu(t)=\mu_a\sin{\omega t}$. The particle-hole amplitudes $u_{\bf k}(t)$ and $v_{\bf k}(t)$ are expected to follow the adiabatic time dependence in Eqs.~\ref{eqsol} with $\mu_0$ continued in time to $\mu_0+\mu(t)$, except at the avoided crossings (marked by arrows) when Landau-Zener transitions are expected.}
\label{avcrossfig}
\end{figure}
\begin{figure}
\resizebox{0.5\columnwidth}{!}{\includegraphics{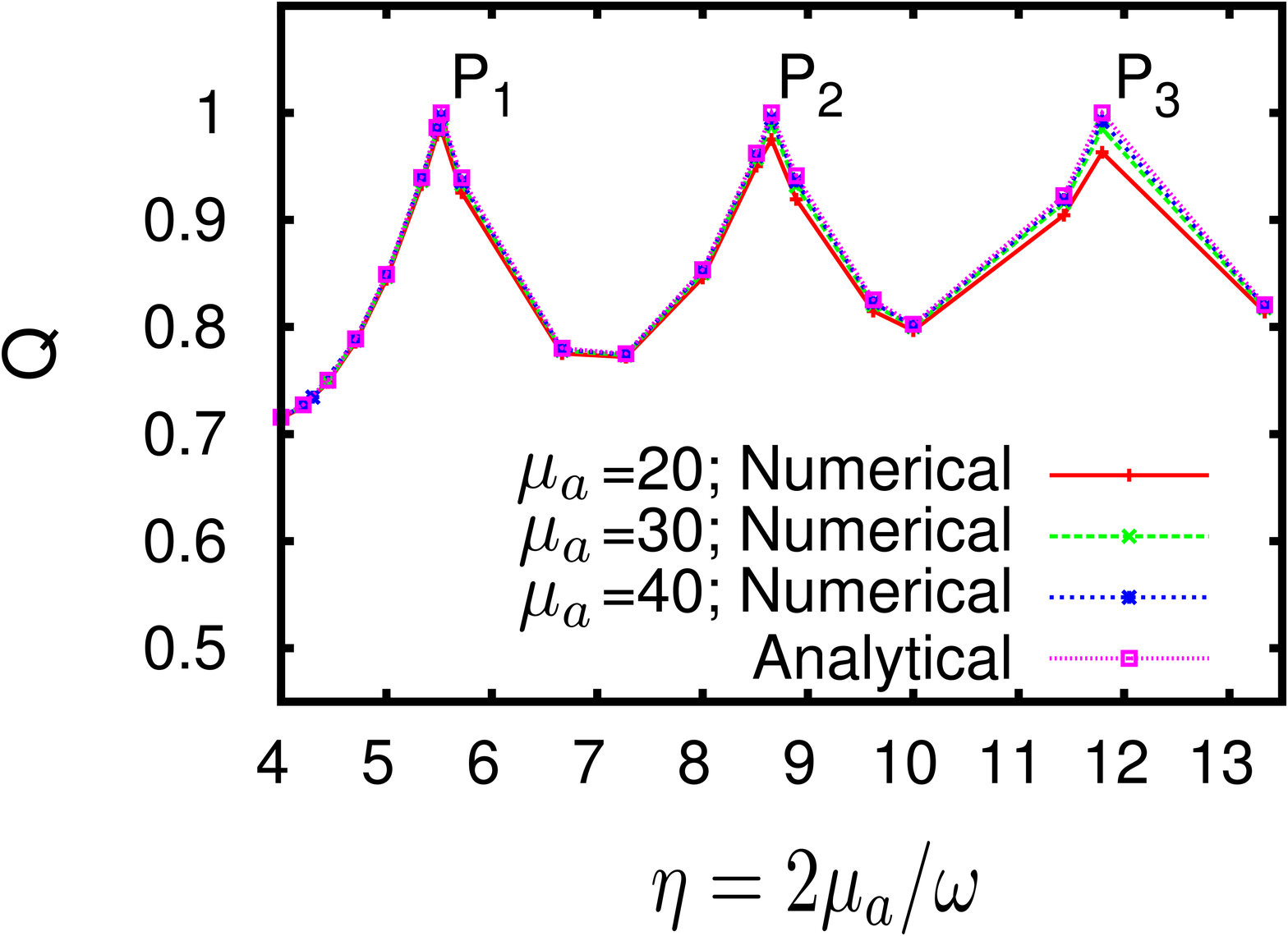} }
\caption{(Color online) Plot of the long time average of the magnetization $Q$ vs  $\eta=2\mu_a/\omega$ for multiple values of $\mu_a$ for $\mathcal{N}=100$ fermions subjected to the non self-consistent dynamics in Eq.~\ref{bdg2}, with $\mu(t)=\mu_a\sin{\omega t}$, and $f_{k}=\cos{k}$, $\Delta(k)=\sin{k}$ for $k\in\{-\pi,\pi\}$ (the Brillouin zone of a 1D optical lattice with unit spacing). These are compared with the analytical result obtained using the Rotating Wave Approximation in~\cite{arnab1}. The peaks $P_{1}$, $P_{2}$
and $P_{3}$ represent maximal freezing, corresponding to three zeros of $J_{0}(\eta)$, occurring at $\eta = 5.520..., 8.653...$, and $11.971...$ respectively. Reproduced from~\cite{arnab1} with permission from author.}
\label{arnabfig}
\end{figure}
 {In this limit, the system stays in the lower of the two adiabatic eigenstates $E({\bf k};t)=-\sqrt{[f_{\bf k}-\mu(t)]^2+\Delta^2({\bf k})}$. More rapid periodicities in a harmonically driven $\mu(t)$ may cause the system to vary across the quantum critical point ($\mu_0=2$), and produce defects away from the adiabatic state~\cite{xyseminal,bikashbabu,review:lzstls}. The density of these defects can be computed for each ${\bf k}$ using Landau Zener St\"uckelberg theory~\cite{landau:lzformula, zener:lzformula} } (reviewed in~\cite{xyseminal,bikashbabu,review:lzstls}) using the adiabatic-impulse approximation. { In this limit, the dynamics of the amplitudes are presumed to be adiabatic for all times except when near an \textit{avoided crossing} in the adiabatic eigenstates \textit{i.e} when the separation between the two eigenstates is at a minimum. This occurs at times $t_c$, given by the solution $f_{\bf k}=\mu(t_c)$ when the energy gap is $2|\Delta({\bf k})|$. The nonadiabatic excitation density for the 
dynamical state as the 
system varies from $t=t_c-$ to $t=t_c+$ was estimated by Landau}~\cite{landau:lzformula} and Zener~\cite{zener:lzformula}  {using the Zener approximation, $\mu(t)\approx \mu_a\omega t$, around the avoided crossing. This process, illustrated by plots of $E({\bf k};t)$ in Fig.}~\ref{avcrossfig} {, yields a defect density of}~\cite{wittig:lzformula}  {$e^{-\chi_0}/2$, a result that is recovered from Eq.}~\ref{eq:lz:pert}  {in the non self-consistent limit. Landau-Zener transition amplitudes over subsequent crossings are modulated by the Stokes phase, a consequence of St\"uckelberg interferometry~\cite{xyseminal, review:lzstls}. A resonance condition in the Stokes phase simplifies the dynamics to one described exclusively by planar pseudospinor rotations, leading to near-harmonic oscillations in the defect density~\cite{xyseminal,rwa}. Even away from such resonances, a continuous Floquet spectrum (obtained in the thermodynamic limit) has been shown to yield periodic behavior~\cite{russomanno}. The Landau-Zener 
transitions also contain classical signatures in the form of the Kibble-Zurek mechanism~\cite{bikashbabu:ref} in the perturbative limit} (see, for instance, Mukherjee and Dutta~\cite{xyseminal}, Damski~\cite{damski}, Dziarmaga~\cite{dziarmaga2}, and, more recently, Russomanno \textit{et. al.}~\cite{russomanno}).  
 
 {Even more rapid values of $\omega$ can produce dynamic symmetry breaking of the time averaged magnetization $Q=\lim_{T\rightarrow\infty}T^{-1}\int^T_0 \mathrm{d}t \quad m(t)$ as reported by Das}~\cite{arnab1} {. This occurs due to the extremely nonadiabatic nature of the dynamics if $\omega\gg\Delta$. This also produces a dynamical hysteresis similar to that from the Landau Zener dynamics. Das}~\cite{arnab1}  {derives the onset of such hysteresis using the Rotating Wave Approximation (henceforth abbreviated as RWA) for large $\omega$ in a manner similar to Kayanuma~\cite{kayanuma:rwa}, Oliver \textit{et. al.}}~\cite{oliver}, as well as Ashhab \textit{et. al.}~\cite{rwa} (also see~\cite{review:lzstls}). Eq.~\ref{bdg2}  {can be approximated by an RWA Hamiltonian that contains only the slowest - oscillating contribution to the time translation operator, given by $\mathcal{U}_{\bf k}(t)=\exp{\{i\tau^z\int^t_0 \mathrm{d}t\quad\left[f_{\bf k}-\mu(t)\right]\}}$ in the interaction picture. This can be obtained by 
a Fourier series decomposition of a periodic $\mathcal{U}_{\bf k}(t)$ and ignoring all the oscillating terms}~\cite{arnab1} {. The resultant time independent RWA Hamiltonian is simply $H^{\rm rwa}_{\bf k}\sim\Delta({\bf k})J_0(\eta)\tau^y$, where $\eta=2\mu_a/\omega$ for amplitude $\mu_a$, and $J_0(x)$ is the zeroth order Bessel function. The Schr\"odinger equation for this Hamiltonian can readily be integrated to yield solutions where the amplitudes $u_{\bf k}$ and $v_{\bf k}$ start to show hysteresis for 
parameters that approach $J_0(\eta)=0$, and dynamically freeze at their equilibrium values at the zeros of $J_0(\eta)$, \textit{i.e} when  $H^{\rm rwa}$ vanishes. The same result can also be obtained from the Floquet theory of time-periodic quantum systems  (see Shirley~\cite{shirley}, Gro\ss mann and H\"anggi~\cite{grossman:hanggi}, Gomez Llorente and Plata~\cite{gomez:floquet}, as well as~\cite{review:lzstls}). This 'exotic freezing' can be seen in the numerical simulations plotted in Fig.}~\ref{arnabfig} { for $100$ fermions in a $1$-D optical lattice of unit spacing. Here, the pairing symmetry is anisotropic, and only the non self-consistent solutions to Eq.}~\ref{bdg2}  {are evaluated. The comparison with the results obtained from the RWA are excellent even at low $\omega$, but are expected to differ when $\omega\lesssim\Delta$, where either the adiabatic or Landau-Zener formalisms yield better results. Unlike the Landau-Zener limit, however, the dynamical hysteresis follows a mechanism that is very 
different from the quantum Kibble-Zurek mechanism}~\cite{arnab1}  {and thus illustrates a uniquely quantum phenomenon.}

\section{\sc Experimental Studies}
The experimental study of nonequilibrium dynamics of many-particle systems using ultracold atoms is a new and exciting field. The tunability of such systems, together with their ability to break new ground in the understanding of many-particle behavior, is unprecedented in condensed matter physics. Experimental studies in this area mushroomed following the realization of the Bose Einstein Condensate in the mid-$1990$s, and have focused on Bose gases (for comprehensive reviews, see~\cite{revs}). Strongly interacting Bose gases are difficult to generate due to three body losses, but much easier in Fermi gases due to the Pauli exclusion principle. This allowed for the experimental realization of the BCS-BEC crossover in Fermi gases using a Feshbach resonance in $2004$~\cite{bcsbec:cross}. In addition, time dependent Feshbach fields have been generated in the laboratory~\cite{cornish} and have been used to study atom-molecule collisions dynamically~\cite{bcsbec:coll}. Impulse quenches of the type discussed in 
this review have also been generated in Bosonic gases~\cite{bikashbabu:ref} These should allow for the study of the dynamics of BCS systems with periodic Feshbach amplitudes, as discussed in this review. In addition, effective time-dependent chemical potentials can be simulated in ultracold atoms by introducing time-dependent harmonic confinements using lasers and acoustic-optical modulators. If the trap is broad enough to allow for local density approximations to the fermion density, then the Bogoliubov de-Gennes  dynamics discussed in this review can be replicated. In addition, real time controls of the periodicity of optical lattices via an 'optical accordion' setup~\cite{hrishi} offer the possibility of periodic drives in BCS systems.

So far, however, most experiments have concentrated on the equilibrium properties of Fermi gases, as well as both equilibrium and nonequilibrium properties of Bose gases. To the author's knowledge, few experiments have been performed on ultracold Fermi gases out of equilibrium till date. Studies of nonequilibrium Fermi gases  range from probing the ballistic expansion of Fermi superfluids~\cite{ohara1}, measurement of their collective excitations~\cite{barnstein1}, and observation of vortex dynamics \cite{zwierlien1}. The diabatic variation of system parameters via quantum quenching in ultracold Fermi gases should allow access to highly coherent and nonequilibrium quantum states of matter hitherto unseen in condensed matter physics. Such quenching has been performed on bosonic $^{85}$Rb atoms using the Feshbach resonance by Donley \textit{et. al.} and others~\cite{bcsbec:coll}. The phenomenon of collapse and revival discussed in the review has been seen in few-body quantum optics systems for some time~\cite{
colrev:oldexp}, and was also seen in collective matter wave packets after a rapid quench past the quantum critical point of the Superfluid-Mott Insulator transition in $2002$~\cite{colrev:exp}. It is also to be noted that the Timmermans Hamiltonian, in the single mode approximation, closely resembles the Dicke model in quantum optics~\cite{yuzbashyan}. Theoretical and numerical work in this area definitely calls for more experiments to be done~\cite{theo:num}. The actual time scales of the ensuing dynamics can be controlled by the system size~\cite{myrefs}. The latter has a wide range of realizable values with ultracold atoms, ranging from a few nanometers using atom chips~\cite{atomchip} to several micrometers using optical lattices~\cite{hrishi}.

\section{\sc Conclusions}
\label{sec:concl}
 {This article has reviewed  numerous works investigating the dynamics of quantum quenching in two distinct regimes of Fermi BCS systems, classified according to the time scales of the quench relative to the equilibrium BCS gap. In one case, the system can be described in the mean field by the time dependent Ginzburg Landau equations of the fermion gap parameter, with a coupling to the Gross Pitaevski Bogoliubov equations for molecular Bosons. Such systems admit to stable solitons, as well as unstable ones in lower dimensions, and impulse quenches away from equilibrium can cause collective behavior that manifest as collapses and revivals of the matter waves. In the other case, the system can be described by self-consistent Bogoliubov de-Gennes equations whose dynamics produce behavior that is markedly different from the non self-consistent dynamics that is appropriate for quenching in Ising and Kitaev models. 

The author thanks CSIR, India for support under the Scientists' Pool Scheme No. $13(8531-$A$)/2011/$Pool. Gratitude is also extended to Prof. K. Sengupta, IACS Kolkata, as well as Dr. A. Das, Max Planck Dresden for their critiques. The author also thanks Prof. G Ambika, IISER Pune and convener of the NCNSD conference, $2012$, for inviting this review, as well as Prof. A. Laxminarayan, IIT Madras, for convening the quantum chaos minisymposium at the NCNSD conference, the proceedings of which form a part of this review.}

\end{document}